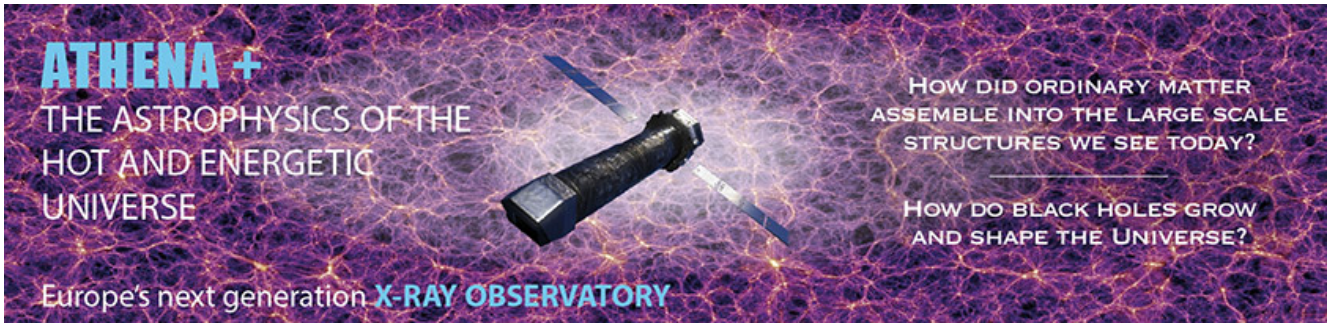

# The Hot and Energetic Universe

An *Athena+* supporting paper

## Astrophysics of feedback in local AGN

Authors and contributors

**M. Cappi, C. Done**, E. Behar, S. Bianchi, V. Braito, E. Costantini, M. Dadina, C. Feruglio, F. Fiore, S. Gallagher, P. Gandhi, N. Grosso, J. Kaastra, A. King, A. Lobban, R. Maiolino, E. Piconcelli, G. Ponti, D. Porquet, K. Pounds, D. Proga, P. Ranalli, J. Reeves, G. Risaliti, P. Rodriguez Hidalgo, E. Rovilos, S. Sim, G. Stewart, F. Tombesi, T.G. Tsuru, S. Vaughan, D. Wang, D. Worrall

# 1. EXECUTIVE SUMMARY

Most galaxies host a supermassive black hole (SMBH) at their center, with a mass tightly correlated with that of the host galaxy (the $M_{BH}$-$\sigma$ relation). These observational facts have revolutionized our view on the formation and evolution of galaxies, and require a self-regulating mechanism connecting the accretion-powered growth of the SMBH to the star-formation-powered growth of the host galaxy. Understanding this *feedback* mechanism is key to understanding the growth and co-evolution of AGN and galaxies (e.g. Fabian 2012 for a recent review). Jet feedback from radiatively inefficient active galactic nuclei (AGN) is well established in clusters of galaxies, where it heats gas in the halo preventing gas cooling onto the central galaxy (radio or maintenance mode, see Croston, Sanders et al., 2013, *Athena+* supporting paper). However, an equally crucial process, in the evolution of massive galaxies, is thought to be the preceding phase in which AGN-driven winds expel gas from the host galaxy, hence quenching star formation. Within this context, the high speed ionized winds recently discovered by *XMM-Newton*, *Chandra* and *Suzaku*, from the radiatively efficient phase of the AGN (Fig. 1, Left), are regarded as the most effective way of transporting energy from the nuclear scale to host galaxy (Fig. 1, Right). According to models, the energy of such powerful AGN-driven winds is deposited into the host galaxy interstellar medium (ISM), resulting into the recently discovered galactic-scale molecular outflows, which are able to sweep away the galaxy's reservoir of gas and quench the star formation activity. The key questions here are:

- How do accretion disks around black holes launch winds/outflows, and how much energy do these carry?
- How are the energy and metals accelerated in winds/outflows transferred and deposited into the circum-galactic medium?

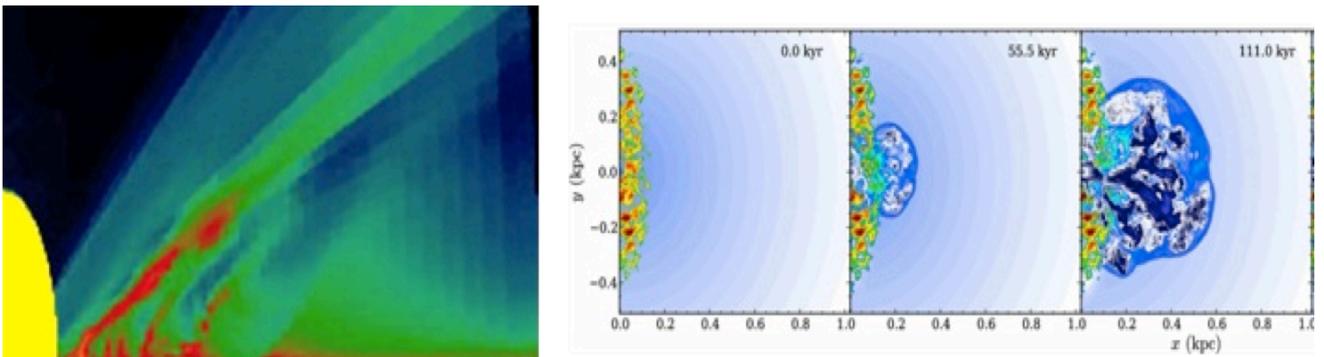

Fig. 1: *(Left)* 2-D theoretical simulation (and movie: http://www.iasfbo.inaf.it/~cappi/athena_plus/qso_2.mpg) of density of an accretion disk wind (red/green/blue) accelerated by UV flux from the disc (bottom) and X-ray radiation from a central source (yellow) (Proga, Stone & Kallman, 2000) *(Right)* Midplane density slices resulting from the "breakout" of a $10^{44}$ erg s$^{-1}$ ultra-fast outflow into a two-phase interstellar medium with clouds distributed in a quasi-Keplerian galactic disk (from Wagner et al. 2013).

X-ray observations are a unique way to address these questions because they probe the phase of the outflow which carries most of the kinetic energy. A high throughput, high spectral resolution instrument is required to determine the **physical parameters** (ionization state, density, temperature, abundances, velocities, geometry, etc.) **on a dynamical time-scale**, in order to assess **the acceleration and launching mechanism of such high speed outflows** by probing the flow close to the SMBH in a broad sample of nearby bright AGN.

The high spatial and spectral resolution of the X-IFU on *Athena+* will also allow direct imaging of the interaction of the wind with the host galaxy for local AGN, forming a template for understanding AGN at higher redshift where wind shocks cannot be resolved.

This combination of spatial and spectroscopic resolution will enhance studies of the closest supermassive black hole, Sgr A*, located at the center of our Galaxy. It is now highly radiatively inefficient, but its rich signs of past and (re)current activity, and interactions with its surrounding environment, make it the nearest laboratory for studying (low-L) AGN feedback.

## 2. HOW DO AGN LAUNCH WINDS/OUTFLOWS AND HOW MUCH KINETIC ENERGY DO THESE WINDS CARRY?

AGN winds are currently directly observed as blue-shifted and broadened absorption lines in the UV and X-ray spectra of a substantial fraction of AGN. These absorption systems span a wide range of velocities and physical conditions (distance, density, ionization state), and each component carries kinetic energy and momentum away from the small-scale AGN into its larger-scale environment. Simulation of accretion disks and outflows have progressed enormously in the last decades, and show that several physical mechanisms (thermal-, radiation, line, or magnetic-pressure) are able to accelerate winds (e.g. Blandford and Payne 1982; Hawley & Balbus 2002; King & Pounds 2003; Proga and Kallman 2004; Ohsuga et al. 2009; Sim et al. 2010; Fukumura et al. 2010, Kazanas et al. 2012), giving a theoretical basis for understanding the observations. But what determines the (energetically) dominant mechanism is not understood.

In the UV and X-ray bands we observe narrow absorption lines outflowing with moderate velocity of hundreds to few thousands km/s. This 'warm absorber' is detected in ~50% of AGN (Blustin et al. et al. 2005; Piconcelli et al. 2005; McKernan et al. 2007), and may have its origin in a swept-up ISM or thermally driven wind from the molecular torus (Blustin et al 2005). The warm absorber carries only a small fraction of the kinetic energy, as the amount of material and outflow velocity are both quite small. Instead, there are two much higher velocity systems which potentially have much greater impact on the host galaxy. In the UV band, broad absorption lines are seen in ~30% of AGN, and may be present but outside the line of sight in most AGN (Ganguly & Brotherton 2008). These absorbers can be outflowing as fast as ~ 0.2c, so carry considerable kinetic energy, and probably arise in a UV line driven wind from the accretion disk (Proga and Kallman 2004).

However, the most powerful outflows appear to be the much more energetic winds so highly ionized that the only bound transitions left are for Hydrogen- and Helium-like iron. Such winds can *only* be detected at X-ray energies, and cannot be accelerated by UV line driving. These X-ray winds are observed in 30-40% of local AGN, with outflow speeds of up to ~0.3c (Tombesi et al. 2010), and in a handful of higher redshift AGN at up to 0.7c (Chartas et al. 2002; Lanzuisi et al. 2012). These 'Ultra-Fast Outflows' (or UFOs) have velocities that point to an origin very close to the SMBH, but the launching and acceleration mechanism remain unclear. Possibilities include radiation-driven winds (when the local flux is above the Eddington limit) and/or magnetic driving, *yet we are still searching for the observational answers to this basic question: how do AGN launch winds/outflows?*

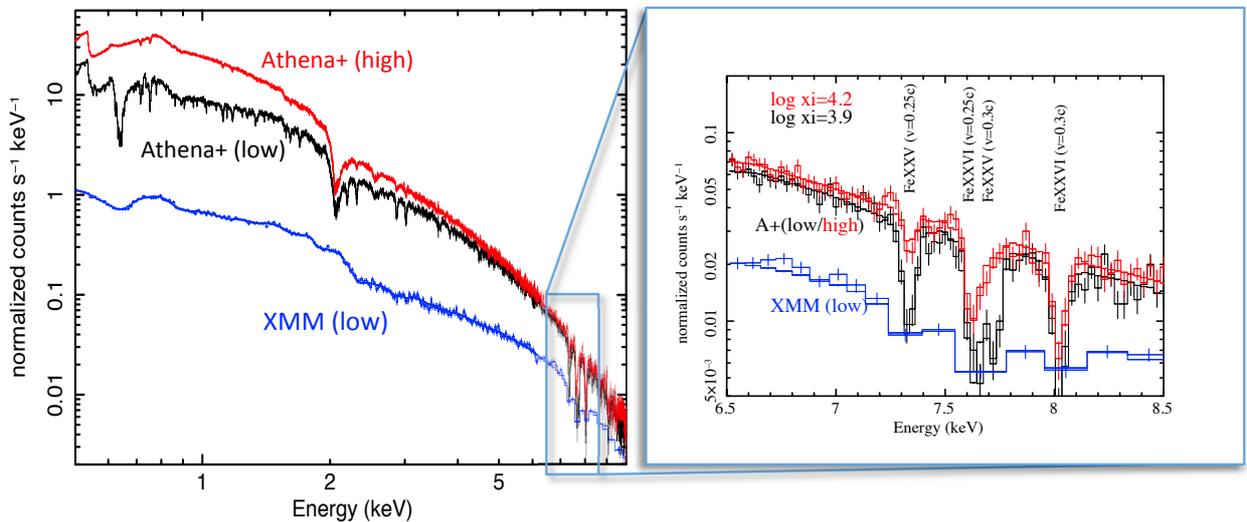

Fig. 2: *(Left) Athena+* X-IFU simulated spectra of the Quasar PDS456 showing the power of resolving simultaneously all ionization lines of the warm absorber component (at soft E<3keV energies), and the UFO component (at higher E>7 keV energies). The spectrum is simulated, for a 100 ks exposure, in two different spectral states (high state in red, low state in black) and illustrates the clear ionization and line variations of the absorber(s) to the continuum flux on short timescales. The *XMM-Newton* spectrum (in blue, for the same low state of *Athena+*) illustrates the greater throughput and energy resolution of *Athena+* w.r.t. *XMM-Newton*. An absorber of size 10 $GM/c^2$ moving at 0.25c would take 200 ks to cross the X-ray source, for a $10^9$ $M_\odot$ black hole. *Athena+* will probe timescales dynamically shorter than this (100ks or less) and resolve the passage of the accretion disc outflow components across the source as well as its location in response to continuum variability. *(Right)* Panel zooming on the variations in the ratios of the He- and H-like Fe absorption lines, for two different velocity systems.

The key to progress on this question is a detailed characterization of the physical properties of these winds (column density, ionization state, outflow velocity, location, geometry, covering factor, etc.) *on a dynamical time scale, comparable to the X-ray source variability time-scale. Athena+* will allow the study of such rapid variability, and seek correlations among the fundamental parameters such as density, ionization parameters as a function of distance from the source, maximum outflow velocity, and X-ray or UV luminosity, which will uniquely constrain predictions of radiation-driven (e.g. Proga & Kallman 2004; Sim et al. 2010), momentum-driven (King 2010), and magnetically-driven (Fukumura et al. 2010) accretion disc wind models. Such high quality data will also quantify the outflow efficiency (i.e. the ratio between the accretion and ejection rates), as well as the kinetic energy budget of the various components in the wind, to better assess their impact on the larger scale environment of the host galaxy (see Section 1.2; Croston, Sanders et al., 2013, *Athena+* supporting paper; and Georgakakis, Carrera et al., 2013, *Athena+* supporting paper).

Figure 2 shows a simulation of the X-ray spectrum resulting from a system showing both forms of X-ray winds: the UFO seen only above 6keV, and the warm absorber producing the many spectral lines seen at lower energies. The calorimeter on-board *Astro-H* will be likely first in probing simultaneously the many low and high energy absorption features from these absorbers, but only for a few of the brightest AGN, and with spectra integrated over observing periods typically much longer than their dynamical time-scales. Only the unprecedented sensitivity of the *Athena+/X-IFU* will allow us to resolve changes in the UFO wind structure for many bright AGN, and on a dynamical (1000s seconds) timescale, hence determine where this material connects to the disk.

*X-ray observations are required to determine the total column density, the highest velocities, the highest ionizations, and hence the total kinetic energy in the winds. Athena+ high spectroscopic throughput will allow for a giant leap of sensitivity to most ionization states of light elements, and to all those of Fe (from I to XXVI), hence providing a detailed characterization and understanding of the AGN outflows on their dynamical timescale.*

## 3. HOW DO AGN WINDS/OUTFLOWS INTERACT WITH THEIR ENVIRONMENT?

As briefly anticipated in the executive summary, understanding disk accretion and ejection processes is fundamental to understanding the origin of feedback and the co-evolution of black holes and their host galaxies. Depending on the covering factors and duty cycles, the mass outflow rate from AGN outflows could reach kinetic powers of a few per cent of the bolometric luminosity, which numerical simulations show exerting a significant impact on the host galaxy (e.g. Hopkins & Elvis 2010; Gaspari et al. 2011a,b; Gaspari, Brighenti & Temi 2012b, Wagner et al. 2013). Therefore, over time, AGN outflows are likely to significantly influence the bulge evolution, star formation, and super-massive black hole growth consistent with the observed ($M_{BH}$– σ) black hole–host galaxy relationship (Silk and Rees, 1998; Fabian 1999; Ferrarese & Merritt 2000; King 2010; Ostriker et al. 2010; Gaspari et al. 2011a,b; Gaspari, Brighenti and Temi 2012; Gaspari, Ruszkowski & Sharma 2012). Other cosmologically important issues in which AGN winds and feedback are likely to play an important role are in: i) limiting the upper mass of galaxies by halting the galaxy growth; ii) providing the extra heating in cooling flows, especially if AGN activity is related to relativistic jets; iii) contributing with "super-winds" from starburst galaxies to the enrichment of the intergalactic medium, and iv) preventing star formation in mergers by effectively removing the coldest ISM gas from galaxies.

While fast AGN outflows can carry a huge amount of mechanical energy from the innermost region nearest the black hole which numerical simulations show will have a significant influence on the host galaxy, how, if, and where, this energy is actually released into the interstellar medium is far from understood. Although there is wide consensus that AGN feedback is likely able to quench star formation, a quantitative understanding of how this process works does not exist (Ciotti, Ostriker and Proga, 2010, and references therein; Ishibashi and Fabian, 2012; Nayakshin & Zubovas 2012; Zubovas et al. 2013; Zubovas and King 2012, 2013; Hopkins 2012; Debuhr, Quataert, & Chung-Pei, 2012; Trump et al. 2013).

Figure 3 shows an example of such a wind-ISM interaction based on recent observations of the narrow line Seyfert 1 galaxy NGC 4051 (Pounds and Vaughan 2011). The *Athena+* simulation illustrates the complex soft X-ray spectra tracing the deceleration of shocked gas following the collision of a highly ionised UFO with the ISM or slower moving ejecta, *at a sufficiently small radius for strong Compton cooling to cause most of the mechanical energy to be lost* (Pounds and King 2013). Importantly, in such a situation, the wind momentum is conserved and provides an alternative feedback mechanism, shown to correctly describe the $M_{BH}$–σ relation (King 2003).

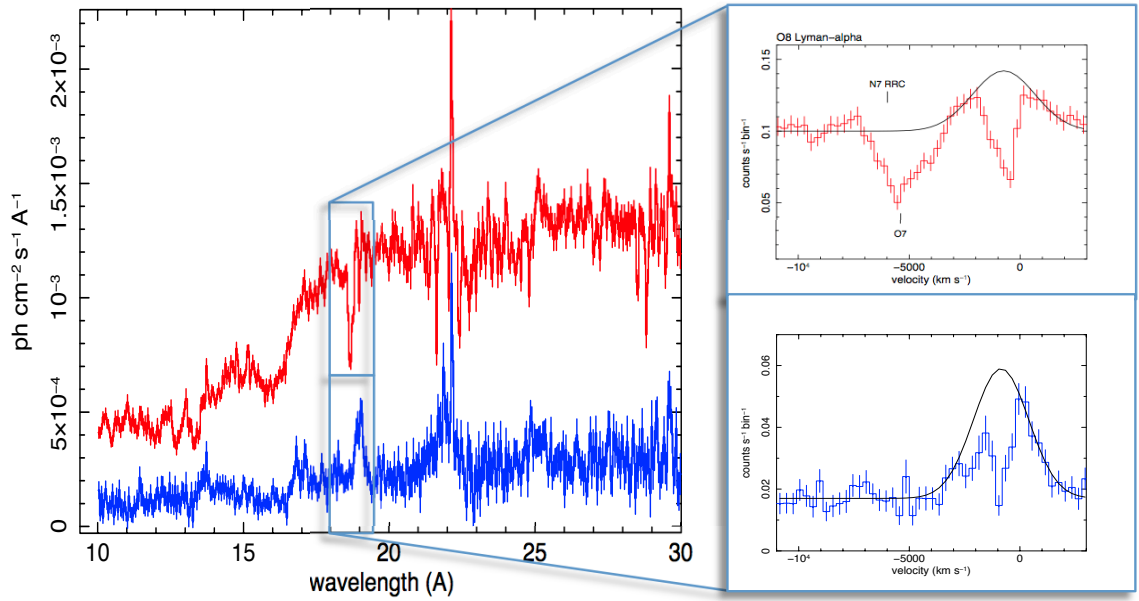

Fig. 3: (Left) Athena+ simulated (~1 ks) soft X-ray spectrum of the Seyfert 1 galaxy NGC 4051 expected from a shocked outflow (as measured by XMM-Newton, Pounds and Vaughan 2011) in an absorption dominated state (top, in red) and in an emission dominated state (bottom, in blue). (Right) The zoomed figures illustrate examples of the richness of spectral details such as blue-shifted broad emission (black line) and self-absorption lines, narrow low velocity absorption, a broad absorption OVIII Lyman trough, NVII radiative recombination continuum, OVII 1s-3p, etc., obtained around the OVIII Lyman alpha spectral feature, which traces the deceleration of the shocked gas, in contrast to the standard view of classical broad absorption line quasars that indicate an accelerating flow.

Observational evidence is mounting for AGN acting on the ISM (molecular and atomic) of their host galaxies on large scales, including the regions of star formation, and, through this, affecting the evolution and the transformation of the host galaxy. Outflows of molecular gas have recently been discovered in both star-forming galaxies and in powerful nearby and distant AGN (Sturm et al. 2011, and references therein), via the detection of both broad emission lines and of P-Cygni absorption lines of several molecular species, at millimeter and far-infrared wavelengths (with IRAM, ALMA, Herschel). The inferred mass outflow rates (several hundreds of solar masses per year) and energetics favour AGN radiation as the driving mechanism, so providing strong support to models of AGN feedback. Evidence of powerful, massive molecular outflows have been also found in a few QSOs at high redshift, including SDSSJ114816, one of the most distant quasars known, at $z$=6.4 (Maiolino et al. 2012, Harrison et al. 2012, Weiss et al. 2012).

In nearby active galaxies, AGN feedback action can be directly probed via detailed imaging and, when possible, via high spatial resolution spectral imaging (i.e. integral field unit, IFU) multiwavelength analysis. IFU data are necessary to allow for a proper 3-D mapping of the various components which overlap spectrally or spatially. Few arcsec resolution is needed to resolve the outflow patterns traced by filaments, bubbles, shells, streamers, etc in nearby active galaxies.

Several studies have been performed in the last decade to trace the larger-scale (i.e. diffuse) soft X-ray emission in dozens of nearby Seyfert galaxies and ultraluminous infrared galaxies (ULIRGs), on scales ranging from a few arcsec to several tens of arcsec (Young et al. 2001; Iwasawa et al. 2003; Huo et al. 2004; Bianchi et al. 2006; Bianchi et al. 2010, Dadina et al. 2010, Wang et al. 2012, 2013; Feruglio et al. 2013). However, to date, determining whether there is a one to one correlation between the soft X-rays and the AGN-excited narrow line region, and whether this extended emission may be related to an outflowing wind (Fisher et al. 2013) has been hampered by insufficient spatial resolution or too low sensitivity. Figures 4 and 5 illustrate the potential capabilities of an X-IFU, as proposed on *Athena+*, for studying AGN-feedback in action in nearby Seyfert galaxies (Fig. 4) and ULIRGs (Fig. 5).

Both AGN and starburst feedback are expected to play important roles in the building of galaxies: mergers will destabilize cold gas and trigger both star formation and nuclear accretion onto super-massive black holes, inducing AGN activity. The ability to separate spatially and spectrally the two types of activity will be key to understand if and how these phenomena are fundamentally linked. The ULIRG NGC 6240 is taken here (Fig. 5) as a local example of an AGN+starburst merging galaxy that should be typical of galaxies at z~1-3, at the cosmic peak of BH and galaxy growth (see Georgakakis, Carrera, et al. 2013, *Athena+* supporting paper). The figure illustrates how an X-IFU will be able to distinguish between starburst-driven excitation (top-right spectrum in Fig. 5) and AGN-heated shocked gas (bottom-right in Fig. 5) as the mechanism responsible for the large-scale galactic outflows of molecular and cold gas.

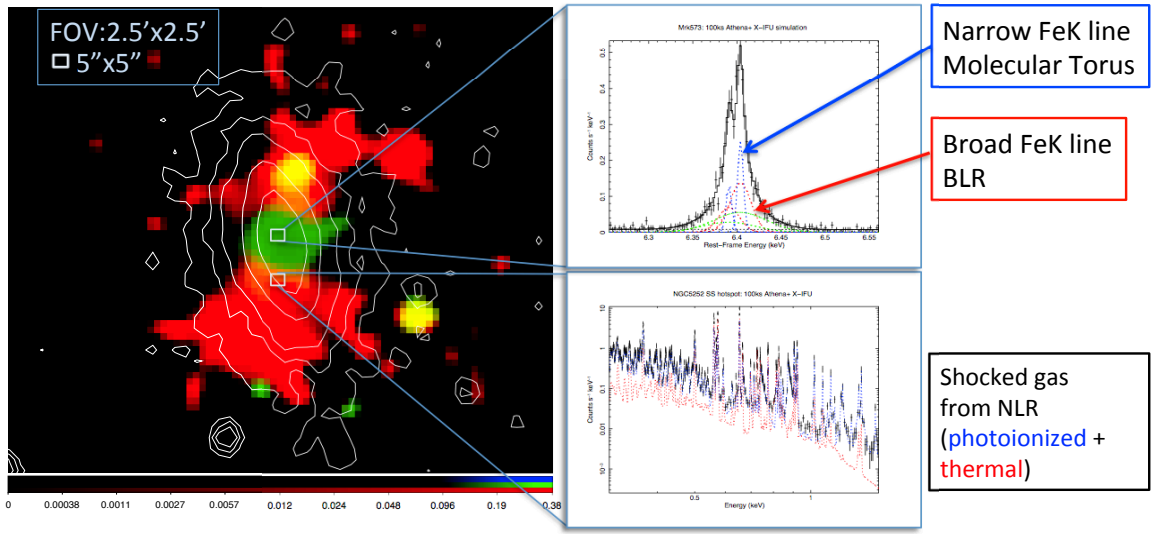

Fig. 4: (Left) Athena+ simulated color-coded image of the nearby Seyfert 2 galaxy NGC 5252 (from the Chandra image in Dadina et al. 2010). The soft X-rays are known (Dadina et al. 2010) to trace the [OIII] ionization cones forming a biconical outflow/illumination pattern driven by the AGN which impacts all over the S0 host galaxy (DSS optical contours indicated in white). (Right-top) Athena+ X-IFU simulated spectrum of the FeK emission line of NGC 5252 assuming the sum of a broad plus a narrow line component, from the BLR and molecular torus respectively (see also Matt, Dovciak, et al., 2013, Athena+ supporting paper). (Right-bottom) Athena+ X-IFU simulated spectrum of a part of the plume of ionized emission south of the nucleus and attributed to 25% of shock (thermal) emission plus 75% of photoionized emission.

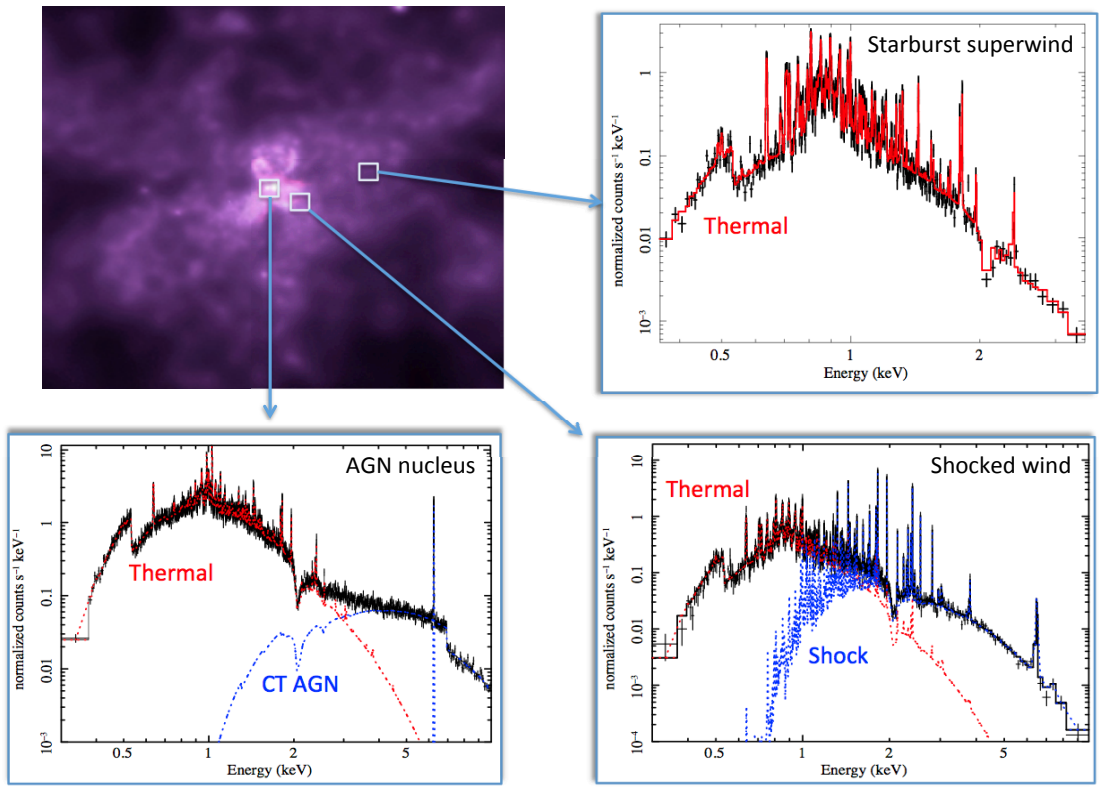

Fig. 5: (Top Left) Chandra X-ray image of the nearby ULIRG NGC 6240 (Nardini et al. 2013). (Top Right) Athena+ X-IFU simulated thermal spectrum from a star-formation driven diffused superwind emission. (Bottom Left) X-IFU simulated spectrum of the nucleus of NGC 6240 including a soft thermal component plus a buried reflected component from the Compton thick AGN double nuclei. (Bottom Right) X-IFU simulated spectrum of part of the plume of ionized emission south of the nucleus and attributed to 25% of starburst superwind (thermal) emission plus 75% of shocked emission. All model parameters are taken from Feruglio et al. 2013 and Nardini et al. 2013.

*For these studies an X-IFU is needed to simultaneously spatially resolve the most prominent components of soft-X-ray diffused emission, and spectrally map the most important nuclear and extended components. The combined spatial and spectral response of the X-IFU on Athena+ will allow astronomers to map the velocity field of the hot gas with uncertainties of ~20-30 km/s on scales down to few kpc in 40-50 nearby AGN/ULIRG/starburst galaxies, thereby allow a direct probe of the interaction of the AGN/starbursts winds with the surroundings in local galaxies and to form a template for understanding AGN-feedback at higher redshift (i.e. SP2.3) where wind shocks cannot be resolved.*

## 4. THE NEAREST LABORATORY FOR "AGN-TYPE" FEEDBACK: THE GALACTIC CENTER AND SGR A* CURRENT AND PAST ACTIVITY

The black hole at the center of our galaxy (Sgr A*) provides a case study in which we can study the interaction (and evidence for previous interaction) between a supermassive black hole and its surrounding environment.

Despite an estimated mass of about $4 \times 10^6$ M$_\odot$, its total luminosity is only 300 times that of the sun, thereby implying both a very low accretion rate and a radiative inefficient accretion flow (see Genzel, Eisenhauer & Gillessen 2010 for a recent review). Sgr A* is typically in a quiescent state, though daily X-ray and NIR flares have been observed, with some periods of increased X-ray activity (Porquet et al. 2008). These flares are intense, reaching tens to hundreds time their quiescent flux on short (hundreds to thousands seconds) time scale. They are believed to originate very close to the SMBH, within few Schwarzschild radii from the horizon (Baganoff et al. 2001, Genzel et al. 2003) and may be evidence of either the sudden increase of the accretion rate, for example due to blobs of material, or the sudden release of magnetic energy and electron accelerations in the flow, or both. To date, the reason for these flares, and in general the accretion and ejection mechanisms at work in the close vicinity of this extremely faint accreting supermassive black hole is not understood.

Current *Chandra* and *XMM-Newton* results indicate non-thermal processes during the flares with relatively soft power-law index for the bright flares (Porquet et al. 2008, Nowak et al. 2012), while for the quiescent state thermal processes seem to take place. However their nature, i.e. collisional or non-equilibrium plasmas is not yet established. Clearly, a high throughput X-ray mission, with good enough angular resolution to isolate Sgr A* from nearby Galactic sources, will allow to follow the source spectral evolution from quiescence to flaring states. Such X-ray studies, in combination with simultaneous observations at other wavelengths (SKA, ALMA, E-ELT), shall allow unprecedented constraints on models for flare production and emission. The spectral resolution of the X-IFU on-board *Athena+* will allow us to discriminate the ionization process for the plasma during the quiescent state thanks to plasma diagnostics, and its physical properties (i.e. Porquet, Dubau, and Grosso 2010). As shown in Figure 6, numerous emission lines will be detected for the quiescent state in a relatively short-time scale (for comparison, 100 times faster than *Chandra*'s HETG) thanks to the X-IFU.

In the last ten years, observations of the Galactic Center have brought numerous new, and unexpected results, in which X-ray observations have given an important contribution. As an example, for the very near future (2013-2014), a cloud of cold gas will undergo a close encounter with the SMBH, and the resulting tidal debris will likely feed the SMBH itself. X-ray monitoring of the evolution of this transient accretion event will likely yield a complementary, important, view of the evolution of Sgr A* accretion state, and possibly accretion disc structure. Other similar events may well happen in the future again, clearly an X-ray calorimeter with sufficient angular resolution to resolve Sgr A* from the diffuse X-ray croud, like *Athena+*, would be mandatory.

Sgr A* is also a case study for understanding SMBH activity in normal (i.e. non active) galaxies (see Ponti et al. 2013 for a recent review). In fact, in the Galactic Center region, we can witness the recent interaction of the SMBH activity with its surrounding environment. In particular there is evidence for Sgr A* to have been activated in the past by the nearby supernova remnant Sgr A East. Another interesting interaction is with the nearby giant molecular clouds (such as Sgr B2, reflection nebulae, etc.). These are reflecting in X-rays the past emission of the central black hole. This has been convincingly shown to be consistent with echoes of activity that Sgr A* underwent a few hundred years ago (Koyama et al. 1996, Ponti et al. 2010, Capelli et al. 2012). If this is indeed the case, Sgr A* was at that time a low luminosity AGN (L$_x$~ $10^{38-39}$ erg/s), making it appearing as one of the brightest X-ray sources in the sky after the Sun.

The Fermi satellite recently discovered large-scale galactic "bubbles" that indicate that in the earlier past (millions of years ago) the source may have experienced a phase of Seyfert-like activity, possibly inflating via an outflow or a relativistic jet (Su et al. 2010, Zubovas et al. 2011). In particular, Zubovas et al. (2011) estimate that, to power the last AGN phase, Sgr A* might have accreted about ~ $2 \times 10^3$ $M_\odot$ that is ten times lower than the estimated mass stored in the circumnuclear disc. The irregular and clumpy structures in the disc suggest dynamical evolution and episodic feeding of gas towards Sgr A* (Liu et al. 2012). The disc appears to be the convergence of the innermost parts of large-scale gas streamers, which are responding to the central gravitational potential well. All these indications suggest that the circumnuclear disc might feed the supermassive BH again (Morris et al. 1999).

Here an X-ray IFU such as the one proposed for *Athena+* will unambiguously characterize the reflections (i.e. echoes) from several X-ray reflection nebulae in the Galactic Center, almost all variable with time. By probing also the faintest such nebulae *Athena+* will be able to trace the activity of Sgr A* over the past millennium, thereby opening a unique window on the study of the past and present activity of a SMBH and its interactions with its host galaxy.

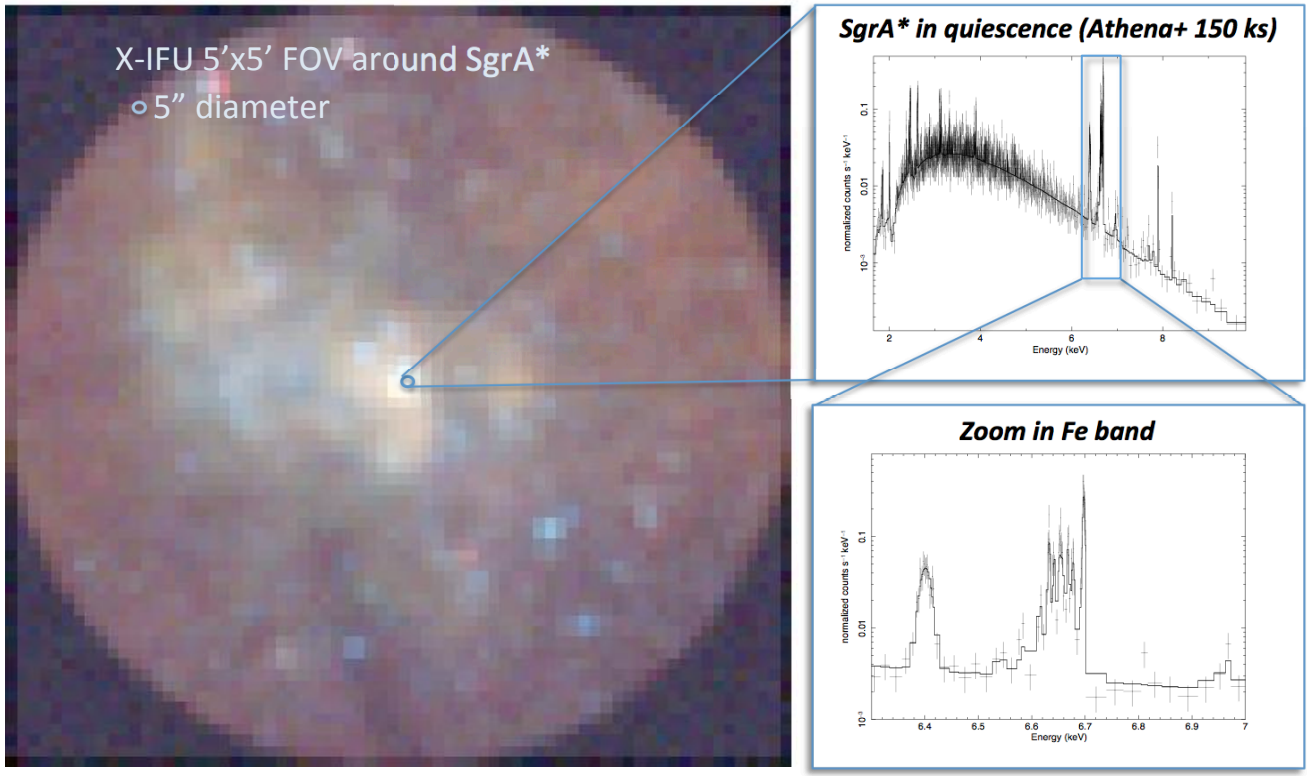

Fig. 6: Athena+ X-IFU (5'x5') simulated figure (with red, green and blue colors coded for 1-3, 3-5, and 5-8 keV energy bands) of the Galactic Center regions with X-IFU simulated spectra (Top right, and zoom bottom-right) of Sgr A* while in its quiescent state (assuming a 150 ks exposure, an absorbed multi-temperature plasma emission with an emission measure following a power-law as in Sazonov et al. 2012). Despite this region being crowded of sources, Athena+ will resolve the brightest off-nuclear X-ray components and isolate Sgr A* emission.